\documentclass[aip,graphicx]{revtex4-1}
\usepackage[a4paper, left=1in, right=1in, top=0.7in, bottom=0.7in]{geometry}
\usepackage{graphicx} 
\usepackage{xcolor}
\usepackage{caption}
\usepackage[label font=bf, 
            labelformat=simple]{subfig}
\usepackage{subcaption}
\usepackage{amsmath}
\usepackage{comment}
\usepackage{svg}
\usepackage{textcomp, gensymb}
\usepackage{upgreek}
\usepackage{xr}
\usepackage{cleveref}

\renewcommand{\vec}[1]{\boldsymbol{#1}}
\newcommand{\mobility}{\vec{\mathcal{M}}}

\newcommand{\kbt}{k_{B} T}

\begin{document}


\title{Emergent clusters in strongly confined systems} 
\externaldocument[S-]{./Supplementary_20251205}


\author{Pamud Akalanka Bethmage}

\affiliation{Department of Physics and Astronomy, Northwestern University, Evanston, IL, USA}

\author{Ryker Fish}
\affiliation{Department of Applied Mathematics and Statistics, Colorado School of Mines, Golden, CO, USA}

\author{Brennan Sprinkle}
\affiliation{Department of Applied Mathematics and Statistics, Colorado School of Mines, Golden, CO, USA}

\author{Michelle M.\ Driscoll}
\email[]{(corresponding author) michelle.driscoll@northwestern.edu}
\affiliation{Department of Physics and Astronomy, Northwestern University, Evanston, IL, USA}


\begin{abstract}
Driven suspensions, where energy is input at a particle scale, are a framework for understanding general principles of out-of-equilibrium organization. A large number of simple interacting units 
can give rise to non-trivial structure  and hierarchy.  Rotationally  driven  colloidal  particles  are  a particularly  nice  model  system  for  exploring  this pattern  formation,  as  the  dominant  interaction between the particles is hydrodynamic.  Here, we use experiments and large-scale simulations to explore how strong confinement alters dynamics and 
emergent structure at the particle scale in these driven suspensions.  Surprisingly, we find that large-scale (many times the  particle  size)  density  fluctuations  emerge  as  a  result  of  confinement,  and  that  these  density fluctuations sensitively depend on the degree of confinement. We extract a characteristic length scale  for  these  fluctuations,  demonstrating  that  the  simulations  quantitatively  reproduce  the experimental pattern. Moreover, we show that these density fluctuations are a result of the large-scale recirculating flow generated by the rotating particles inside a sealed chamber.  This surprising result shows that even when system boundaries are far away, they can cause qualitative changes to mesoscale structure and ordering.


\end{abstract}


\maketitle
\section{Introduction}

Suspensions of driven or active particles are well-known to organize into striking patterns and formations~\cite{marchetti2013hydrodynamics,ramaswamy2010mechanics,gao25}, and this emergent structure often provides a window into the complex interactions that control these out-of-equilibrium systems.  In many, if not most, active matter systems (colloids, bacteria, etc.) hydrodynamic interactions couple the kinematics of the individual agents, profoundly influencing collective dynamics and the emergence of structure~\cite{marchetti2013hydrodynamics, driscoll19, Lushi2014, TangSwarm2021, Yeo2015}. Understanding how these interactions are modulated by confinement, background flow, body shape, or interfacial forces is essential for deciphering how local dynamics give rise to long-range order. Without a detailed grasp of the hydrodynamic coupling mechanisms at play, efforts to predict, control, or exploit emergent behavior in driven and active and matter will remain incomplete.  


A growing body of research has demonstrated that collective interactions in driven and active colloidal suspensions can lead to spontaneous cluster formation and significant particle density fluctuations~\cite{vlahovska2019, bartolo2013,igor2017,snezhko2020}. In particular, driven colloids near planar boundaries have been shown to self-organize into clusters with a characteristic wavelength~\cite{driscoll17}; sedimenting colloidal monolayers on inclined planes form sawtooth wavefronts~\cite{sprinkle21}; and active magnetic colloids confined in quasi-2D wells can spontaneously form vortices with tunable functionalities~\cite{snezhko2024}. These large-scale, sustained (particle) density anomalies are often triggered by hydrodynamic instabilities~\cite{delmotte17,delmotte_minimal_model} or emerge through particle interactions such as Motility-Induced Phase Separation~\cite{MIPS_Quincke_Bartolo}. However, the effects of confining geometry on the density of driven particle suspensions has not been well characterized. We show that even in initially homogeneous systems, and in the absence of geometric structure (e.g., arrays of pillars), confinement alone can induce persistent, long-wavelength density fluctuations.

In our work, we study structure formation and transport dynamics in a sedimented dense suspension of microrollers, rotationally driven colloidal particles.
Using both experiments and large-scale simulations, we study how structure formation is altered when the system is strongly confined, e.g.\ confinement on the same order as the particle length scale.  We demonstrate that full confinement (no-slip cuboidal domain) strongly modifies particle transport and qualitatively changes the suspension structure: large-scale, transient density fluctuations appear in strongly confined systems.  Using high-resolution flow-field calculations, we show that these density fluctuations arise from large-scale recirculation caused by distant lateral boundaries (thousands of particle diameters away) which qualitatively alter the suspension structure. These boundaries are ever-present, as nearly all experimental systems are contained inside a sealed chamber. These results demonstrate that remote boundaries can fundamentally alter suspension configuration, even in a system with strong  confinement where hydrodynamic screening might be expected.

\section{Methods}

We use both experiments employing microrollers and large-scale simulations to study driven suspensions in strong confinement.

\subsection{Numerical Methods }
\label{sec:numerical_methods}
We simulate particle dynamics at the Stokes regime as
\begin{equation}
    \vec{U} = \mobility \vec{F}
\end{equation}
where $\mobility$ is the particle mobility matrix, $\vec{F}$ is a combined vector of forces and torques on the particles, and $\vec{U}$ is the resulting linear particle velocities. The mobility matrix encodes the many-body hydrodynamic interactions between particles as well as the effect of confining boundaries on the hydrodynamics. We compute the action of $\mobility$ using the DPStokes method from the GPU-accelerated library \texttt{libMobility}, which is a force-coupling method using non-Gaussian kernels \cite{hashemi23, libmobility, FCM_2010}. Particle positions are integrated in time using second-order Adams-Bashforth,
\begin{equation}
    \vec{x}^{n+1} = \vec{x}^n + \Delta t \left[ \frac{3}{2}\mobility^{n} \vec{F}^n - \frac{1}{2} \mobility^{n-1}\vec{F}^{n-1} \right].
\end{equation}
We used a timestep of $\Delta t = 5 \times 10^{-4}$ s.
The boundary conditions from DPStokes only include vertical confinement as the method is otherwise doubly-periodic in the $xy$-plane. To incorporate the effects of lateral confinement into the simulations, we build walls near the edges of the periodic domain using particles fixed in place with springs with stiffness $\kappa = 5$ aJ. Here, the domain size is either $500a \times 500a \times H$ or $1000a \times 1000a \times H$ and $H$ is varied across simulations; see SI section V for additional details regarding the design of no-slip walls. Driven particles of radius $a=1.02$ $\mu$m (to match the experimental particles) are initialized randomly (but non-overlapping) in the plane, and with heights sampled uniformly from $z \in \left[a, 2a\right]$ at an area packing fraction of $\phi=0.32$. To drive the particles, we apply a constant torque of $\vec{T} = 8 \pi \eta a^3 \Omega \hat{y}$, with $\Omega = 2\pi f$ and $f=9$ Hz. Contrasting with experiments, the applied torque does not alternate directions (in the experiments the $\hat{x}$ direction of the field alternates signs every 60 s to avoid particle accumulation at boundaries), and  only particles in the interior $50-70\%$ of the domain are used in the data analysis to avoid including the effects of pile-ups at the edges of the domain. To prevent particles from overlapping with each other or the confining surfaces, we include a purely repulsive potential of the form~\cite{floren_blaise_rollers17}

\begin{equation}
    U(r) = U_0 \begin{cases}
        1 + \frac{d-r}{\lambda} & r < d, \\
        \exp \big(-\frac{r-d}{\lambda}\big) & r \geq d.
    \end{cases}
\end{equation}
The parameter $d$ is based on when a particle overlaps another object. For particle-particle interactions we take $d=2a$, and for particle-wall interactions we take $d=a$. We use $U_0 = 8\kbt$ at $295$ K and $\lambda = 0.02a$. We also include the particle buoyant weight as $mg=0.0303$ pN and set the viscosity of the fluid to $\eta=0.9544 \times 10^{-3}$ Pa s to match the experiments.  All simulation parameters are listed in SI Section I.

\subsection{Experimental Methods}
\label{sec:exp_methods}

Here we employ the microroller system that we have used in previous works~\cite{driscoll17,sprinkle20,driscoll2023}.
Microrollers are core-shell colloidal particles composed of a hematite (Fe$_2$O$_3$) core enclosed in a polymer (3-(trimethoxysilyl)propyl methacrylate) shell. Our suspension is highly monodispersed, with a particle diameter of $ 2.03 \pm 0.04$ $\upmu$m and a cube side length of $0.77 \pm 0.1 \: \upmu m$ ~\cite{sprinkle20}; an SEM of the particles is shown in the inset of Fig.~\ref{fig:colloid_config}a.  The particles are driven via an applied magnetic field.  We note that the dominant particle-particle interaction in this system is hydrodynamic, as the hematite  dipole moment  ($|m| \sim 5 \times 10^{-16}$ A m$^{-2}$)~\cite{sprinkle20,driscoll17} is strong enough to couple an external field, but weak enough such that any dipole-dipole interactions are negligible compared to thermal ($ < 0.1 k_{B}T$) and viscous stress ($ Ma \approx 500 $) ~\cite{sprinkle20, driscoll17,delmotte17}. A rotating magnetic field $\vec{B}$ applies a torque $ \vec{\tau} = \vec{m} \times \vec{B}$ to the microrollers; if this torque is greater than the viscous torque on the particle, the particle will rotate synchronously with the field.  Following previous work~\cite{sprinkle20}, we ensure this condition is met by operating at a field amplitude of 85 G and and a frequency of 9 Hz.  For a given area fraction, the translational velocity of the driven suspension increases linearly with both the frequency (up to a cutoff frequency where the magnetic torque can no longer overcome viscous torque) and the suspension area fraction.~\cite{driscoll17} 

The rotating magnetic field ($|B| = 85$ G, $f = 9$ Hz) in the experiments is always in the  $x-z$ plane, and  is generated using computer-generated signals driven through current amplifiers and then fed into triaxial Helmholtz coils. The Helmholtz coils are mounted on top of an inverted fluorescent microscope (IX83, Olympus) so that the sample can be actuated and imaged simultaneously~\cite{sprinkle20}.

\subsubsection{Experiments with the roller transport (suspension velocity measurements)}

To measure the suspension velocity, we fluorescently labeled a small number of particles to allow us to perform single particle tracking in a dense suspension.  
A mixture of otherwise identical fluorescent and non-fluorescent microroller particles (from the same synthesis batch) were used, with the concentration adjusted so that a small amount of particles were florescent labeled (roughly 5-10 per field of view). For these experiments, commercial rectangular capillary tubes were employed (Vitrocom).  Microrollers were suspensed in a  0.14mM KCl solution and  sealed using a UV glue (Noland No.\ 68). Velocity measurements were made by using particle tracking algorithms (Trackpy~\cite{allan2025}) to locate the fluorescent particle positions, and then compute instantaneous velocities (velocity average over 200 ms). We also use standard PIV tools~\cite{piv}  to obtain the 2D (e.g.\ projected) velocity fields. The direction of the driving field was flipped every 60 s for the $h^* =100, l^* =1000$ and $h^* = 40, l^* = 400$ channels and every 20 s for the $h^* = 10, l^* = 100$ channel. The data was sliced to account for the flipping before the velocities were computed. For these measurements, we maintain an area fraction $\phi \sim 0.5$     

\subsubsection{Highly confined samples (suspension structure measurements)}

To achieve the large confinement in $\hat{z}$ without strong confinement in $\hat{y}$ was not possible with commercially available chambers 
Thus, to make large (in-plane) samples with $h^* = 10$, we fabricated custom chambers.  We cut a standard cover-slip (Fisherbrand) using a diamond cutter to get a glass piece roughly $7$ mm $\times 7$ mm in area. This piece of glass along with a standard microscope slide is washed using IPA and dried using a nitrogen air gun. To form the gap between the cover-slip and the slide, a very dilute concentration of   polystyrene spheres (Spherotec FP-10052-2 Flourescent Yellow, diameter $10.6 \upmu$m) is first pipetted onto the microscope slide and sandwiched with the glass piece. To ensure the particles and chamber are in physical contact, we then place the assembled chamber in an oven at $60 \degree$ C for 10--15 minutes; solvent evaporation seals the sample together. Then, microrollers are suspended in a $0.1\%$ w/w  solution of F108 surfactant ($\phi \sim 0.32$). This solution is then wicked into the constructed chamber and then the chamber is sealed using UV glue (Noland Optical No.\ 68).  To verify the gap height in the constructed chamber, we separately coat each glass piece (coverslip and slide base) with 1 $\upmu$m particles with an excitation wavelength in a different channel than the microrollers.  By focusing separately on each plane of particles, we can use the translation of the microscope objective to measure the gap between the two glass surfaces. The $\hat{x}$ direction of the field (e.g., the driving direction) was alternated in sign every 60 s to avoid particle accumulation at the sample edges. Images were acquired at a frame rate of 2 fps.

\subsection{Spectral analysis to identify pattern length scale}

We use spectral analysis (Fourier transforms of images) to characterize the length scale of the emergent pattern.  To ensure the simulation and experimental data is processed in the same manner, we create a 2D projection image from the simulation particle positions to mimic the two dimensional projection of the microscope's field of view. The pixel resolution in the experimental image is $ \Delta x = 0.293 \: \upmu m$, which forms the upper-bound of the spatial frequency range, and the lowest spatial frequency we can measure is limited by our field of view of $\sim 350a \times 350a$. In the simulations, we use a field of view of $ \sim 350a \times 350a$ or $ \sim 500a \times 500a$ depending on whether we use a grid size of $l^* = 500$ or $l^* = 1000$. We perform thresholding in all images, as it is necessary to reduce the background noise in the experimental data. Before taking the Fourier transforms of the projected images, we use a Hanning window to avoid artifacts due to edge effects. 
 
The intensity fluctuations \( \delta\psi(x,y) = \psi(x,y) - \langle \psi(x,y) \rangle\) of the 2D particle distributions is then used to obtain the absolute values of the amplitudes and subsequently the power spectral density $S(u_x,u_y)$ = \( |\Psi(u_x,u_y)|^2\) for both the simulation and experimental data. The transition from a randomly distributed particle configuration to visible clusters generates a low frequency ring in the power spectra. Due to azimuthal symmetry, we perform a radial average to get a one dimensional power spectrum $S(u_r)$ where $u_r = \sqrt{u_x^2 + u_y^2 }$.

We temporally average our data once the steady state pattern appears after a time scale $t > t_{tr}$, see SI section II for details regarding the time scale characterization. We normalize the radially and temporally averaged power spectra with a power spectrum  averaged over particle initial configuration. The power spectrum we use for normalization is representative of the initial condition of the particle distribution (e.g., when the particles are uniformly distributed). This normalization aids in suppressing artifacts near the center of the spectrum which can arise due to spectral leakage.

\subsection{Number fluctuations to quantify clustering}

\label{sec:number_flucs}
To further characterize the clustering pattern, we measure number fluctuations.  This measurement can only be made with the simulation data, as we do not have position data for all particles in the experimental system.  We use particle number fluctuations both to quantify the clusters as a function of time in the larger grid $l^* = 1000$, and to quantify the clusters as a function of $h^*$ in the $l^*=500$ grids. As with the spectral measurements, we focus only on the particles on an interior portion of the grid, away from the fixed lateral walls. The size of this window is $ 350 \times 350 \: \upmu m$  for $l^{*} = 500$ and $ 500 \times 500 \: \upmu m$  for $l^{*} = 1000$. This portion of the domain is then subdivided into bins of size varying from $3.5a$ to $35a $ (for $l^{*} = 500$)  and $ 3.3a $ to $ 50a $ (for  $l^{*} = 1000$). The expected number of particles in each bin is computed ($\bar{N}$) along with the standard deviation of the particles ($\sigma$) about this mean value ($\bar{N}$). We then plot $\sigma$ vs. $\bar{N}$ for different instants of time and as a function of $h^*$.

\subsection{Computing spatial velocity correlations}

\label{sec:correlation}
We characterize the dynamics of the emergent pattern by computing correlations in suspension velocity.
We use standard PIV tools \cite{piv} to calculate the temporally averaged velocity fields $\vec{V} = (V_x, V_y)$ for both experimental and simulation images. This time averaged velocity field is used to extract the fluctuations about the mean velocity $ \delta \vec{V} = \vec{V} - \langle \vec{V} \rangle$. We then compute the 2D autocorrelation of the velocity fluctuations $\delta \vec{V}$ using standard packages in python to obtain $\tilde C_{space}(u,w) = \sum_{i j}  (\delta V_x(i+u, j+w).V_x(i,j) + V_y(i+u,j+w).V_y(i,j)) $; $u,w$ here are lag spacings in $\vec{i}, \vec{j}$ directions. We perform an average over the azimuthal direction in polar coordinates to obtain $C_{space}(r)$ where $r = \sqrt{u^2 + w^2 }$, which is normalized by the combined variance $ C(0) =  \langle |\delta \vec{V}|^2 \rangle $.  A fit to an exponential decay is then used to extract a correlation length.

 \section{Results}

\begin{figure}[htbp]
    \centering
\includegraphics[width=0.9\textwidth]{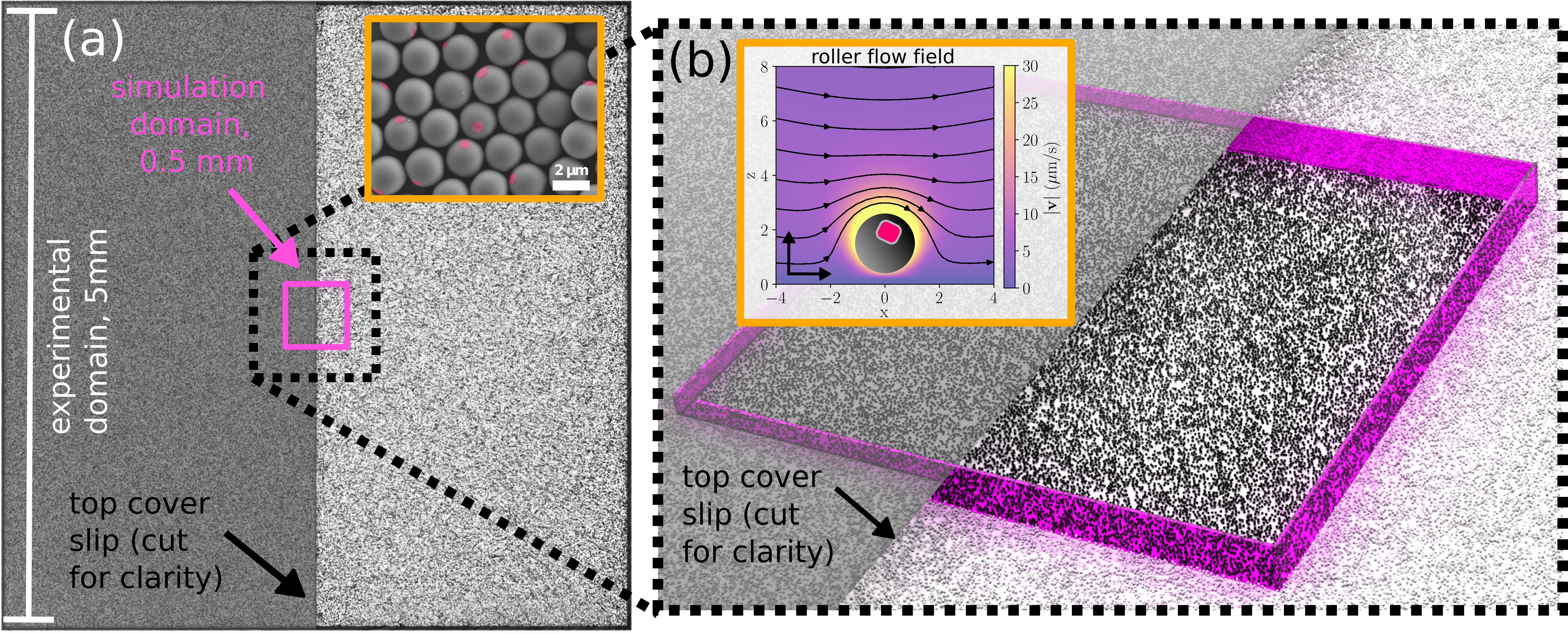}
  \caption{
  (a) Illustration of a roller suspension $(\phi=0.32)$ at the experimental scale with the pink square showing the smaller scale used in simulations. The domain is strongly confined vertically by a bottom and top cover slip. Inset: SEM image of the microrollers; magnetic core is highlighted in red. (b) Magnification of the smaller simulation domain. Pink walls indicate location of lateral confinement in the simulation. Inset: high-resolution simulation of the velocity field generated by a microroller.}
    \label{fig:colloid_config}
\end{figure}

In this work we explicitly look at the case where a driven suspension is confined between in a narrow channel using experimental and simulation tools. 
In the experimental system, we use weakly magnetic core-shell colloidal particles as illustrated in Figure~\ref{fig:colloid_config} (see more details in Methods). Here, the particle magnetic moment is large enough to couple to the external field and allow rotational driving, but small enough so that interparticle magnetic interactions are negligible compared to thermal and viscous stress~\cite{sprinkle20, driscoll17}. The dominant interparticle interaction is hydrodynamic, and suspension structure and dynamics can be quantitatively captured by hydrodynamic simulations~\cite{driscoll17, delmotte17, sprinkle20, driscoll2023}. 

Our simulations include far-field interactions between particles and system boundaries, as described in more detail in Methods. We drive the particles by adding a constant torque corresponding to the same angular frequency applied in the experiments, and use a GPU-accelerated version of the force-coupling method to allow efficient calculation in a large domain~\cite{libmobility, FCM_2010, hashemi23}.

Due to the magnetic core, the microrollers are heavy and sediment near to the chamber floor, but their small size means thermal fluctuations act to keep them at an average height $h_g$. 
When a rotating magnetic field is applied in the $x-z$ plane, the microrollers synchronously rotate with it, providing the field amplitude provides sufficient torque~\cite{driscoll17}. This driving leads to translation due to hydrodynamic coupling with the chamber floor, as illustrated in Fig.~\ref{fig:colloid_config}.  In a dense suspension of microrollers, our previous work has shown that interparticle hydrodynamic interactions generate a large array of emergent structure such as layering, shock formation, and fingering patterns~\cite{driscoll17,sprinkle20,delmotte17}.  Here, we explore how suspension transport properties and structure change with particle-scale confinement, and demonstrate that large-scale pattern formation is induced by confinement. We show that the key driving mechanism for this pattern formation is the microroller-generated recirculating flow which results from the suspension being confined not only vertically, but laterally as well, e.g.\ the experiments are always done in a sealed chamber.



\begin{figure}[htbp]
    \centering
         \includegraphics[width=0.8\textwidth]{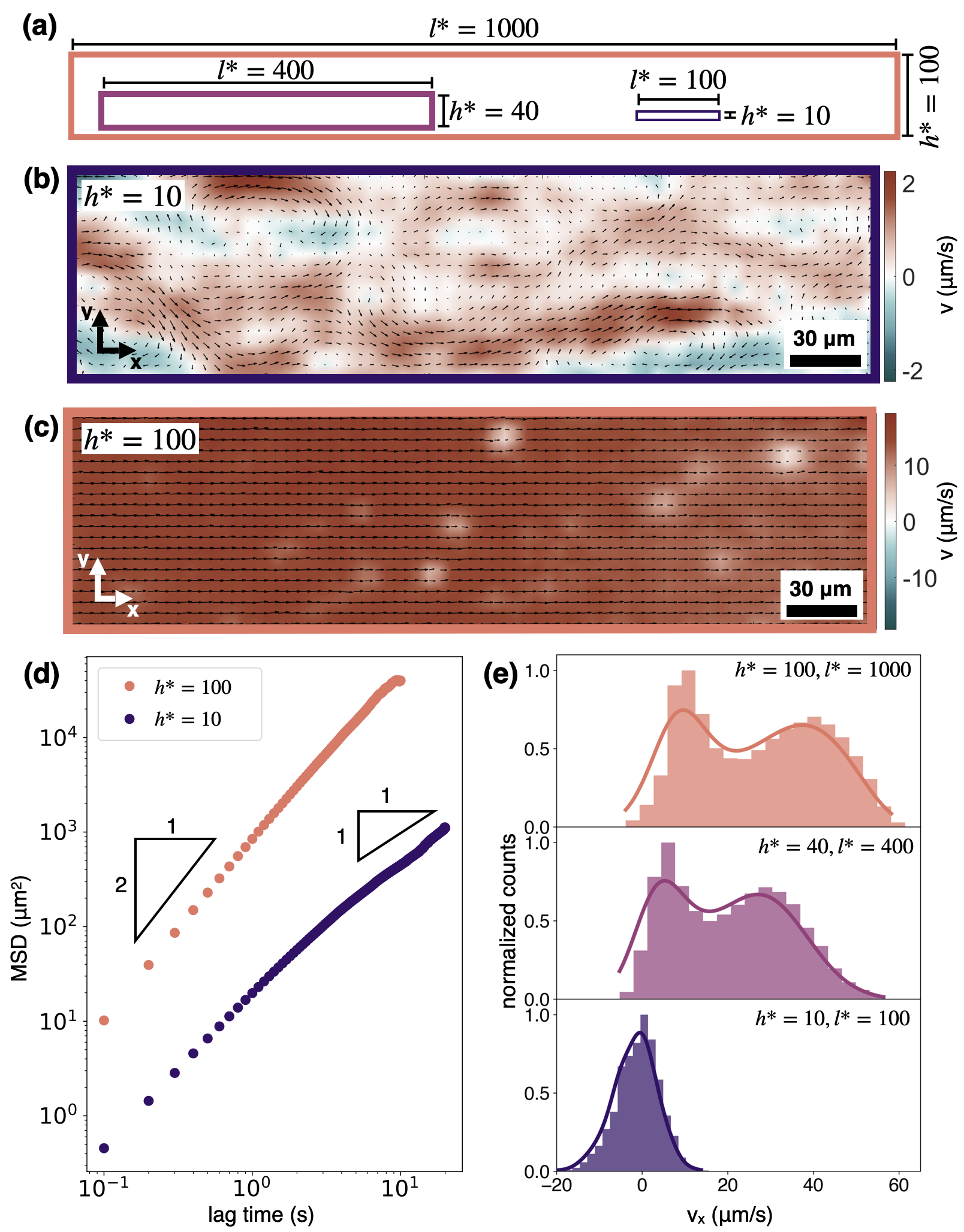}
    \caption{ (a) Cross-sectional geometry of the commercial channels used (all lengths are in dimensionless units of particle radii). (b,c): PIV measurements of the flow velocity in channels of (b) $h^* =10$ and (c) $h^* = 100$.  Multimedia available online: brightfield image corresponding to (b) (d) MSD of the particles in (b) and (c). (e)~Experimental measurements of the suspension velocity distribution, (top) $h^*=100, l^*=1000$ (middle)  $h^*=40, l^*=400$ and (bottom) $h^*=10, l^*=100$.}

    \label{fig:velocity_distributions}
\end{figure}

\subsection{Confinement alters suspension dynamics. }

We measure the transport properties of uniform suspensions of microrollers in channels with varying amounts of lateral, $ l^* = l/a $, and vertical confinement, $h^* = h/a$, where $h$ is the height of the chamber (in $\hat{z}$), $l$ is the width of the chamber (orthogonal to driving, in $\hat{y}$), and $a$ is the radius of the microroller (see Fig.~\ref{fig:velocity_distributions}a). Our previous work demonstrated that when the suspension is not confined vertically $h \gg a$ (e.g.\ $h^* \rightarrow \infty$), the suspension separates into two layers, a faster layer advected over a slower layer, so that the velocity distribution in the suspension becomes bimodal~\cite{sprinkle20}.  Our more recent numerical studies demonstrate that as the suspension is confined, this effect is suppressed as $h^*$ is decreased~\cite{hashemi23}.  The bimodal distribution becomes unimodal and the mean suspension velocity is greatly reduced.  Here, we examine in an experimental system how the particle motion is altered as a function of $l^*$ and $h^*$.

 We measure suspension velocity by tracking the trajectories of individual microrollers using particle tracking~\cite{allan2025}.  To enable measurement in high area fraction suspensions ($\phi \sim 0.5$), we dope a small fraction of fluorescent microrollers into a suspension of largely non-fluorescent particles~\cite{sprinkle20}. We observe that as confinement is increased, the suspension no longer moves in a uniform direction, but instead observe particles moving erratically and even in the direction opposite the driving direction ((Multimedia available online, Figure 2 Video).  This is clearly seen when looking at PIV measurements of the suspensions (Fig.~\ref{fig:velocity_distributions}b,c): at high $h^*$ the suspension is uniform, but at low $h^*$ it displays large inhomogeneities in velocity. This is confirmed by measuring particle mean squared displacement (MSD) as shown in Fig.~\ref{fig:velocity_distributions}d: at large $h^*$ it is ballistic (MSD$ \sim t^2$), but at small $h^*$ it is diffusive (MSD$ \sim t$) at higher lag times, reflecting the disordered motion. Additionally, we observe that the bimodal suspension velocity  distribution observed at low confinement (large $h^*$) becomes unimodal and has a mean near zero at strong confinement (small $h^*$), as shown in Fig.~\ref{fig:velocity_distributions}e.  Due to our use of commercial channels, we do not here independently vary $h^*$ and $l^*$; however we believe the dominant changes in the confinement dynamics are a result of vertical confinement (e.g., in $\hat{z}$).  As we show in the following section, pure vertical confinement strongly alters suspension dynamics even when $l^* \gg 1$, likely because the flow field generated by a roller extends in the $x-z$ plane and decays faster in the $y$ dimension in the $x-y$ plane.

\subsection{Confinement alters suspension structure. }

\begin{figure}[htbp]
       \includegraphics[width=0.9\textwidth]{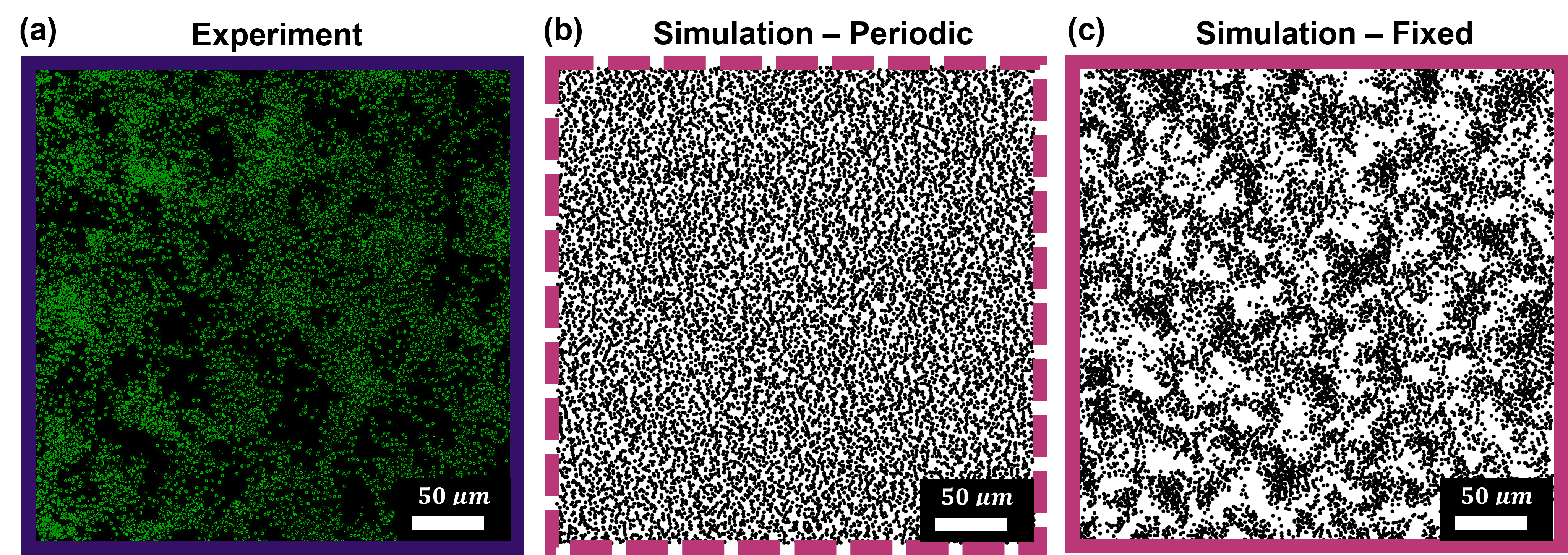}
        
        \caption{ (a) Experimental image showing the pattern formation we observe in a highly confined sample ($h^* = 10,$ $\phi = 0.32$); image shows the center of the channel. (Multimedia available online) (b,c) Numerical simulations ($l^*=1000$) of the experimental system with (b) periodic boundaries and (c) fixed lateral boundaries (as illustrated in Fig.~\ref{fig:colloid_config}). (Multimedia corresponding to (c) available online).  All images show the system after it has evolved to a steady state starting from an initially uniform distribution.}
        \label{fig:density_fluctuations}
\end{figure}

In addition to modifications in suspension velocity, vertical confinement creates large changes in suspension structure: the initially homogeneous suspension exhibits large-scale density fluctuations, see Figure \ref{fig:density_fluctuations} (Multimedia available online: Figure 3a Video).  To alter $h^*$ without modifying $l^*$, we switch from commercial channels to custom-fabricated channels. 
As shown in Figure \ref{fig:density_fluctuations}a, large-scale density fluctuations (e.g., patterning) are seen in the experimental system (we note the view shown is not the whole chamber, but is a view from the center of the channel). To further explore how $h^*$ controls the patterning scale we simulate the system using a force-coupling method.

When these lateral walls are added to the simulation, we reproduce the patterning observed in the experiment, see Figure \ref{fig:density_fluctuations}c (Multimedia available online, Figure 3c Video).  This provides strong evidence that the patterning we observe is a result of the large-scale flow set up in the doubly confined chamber by the microrollers (see \Cref{disc}).  We further note that these simulations were deterministic, and did not include thermal fluctuations, indicating that stochasticity does not appear play a role in the observed density fluctuations and that the patterning we observe can likely be solely attributed to far-field hydrodynamics.

\begin{figure}[htbp]
    \centering
\includegraphics[width=1.0\textwidth]{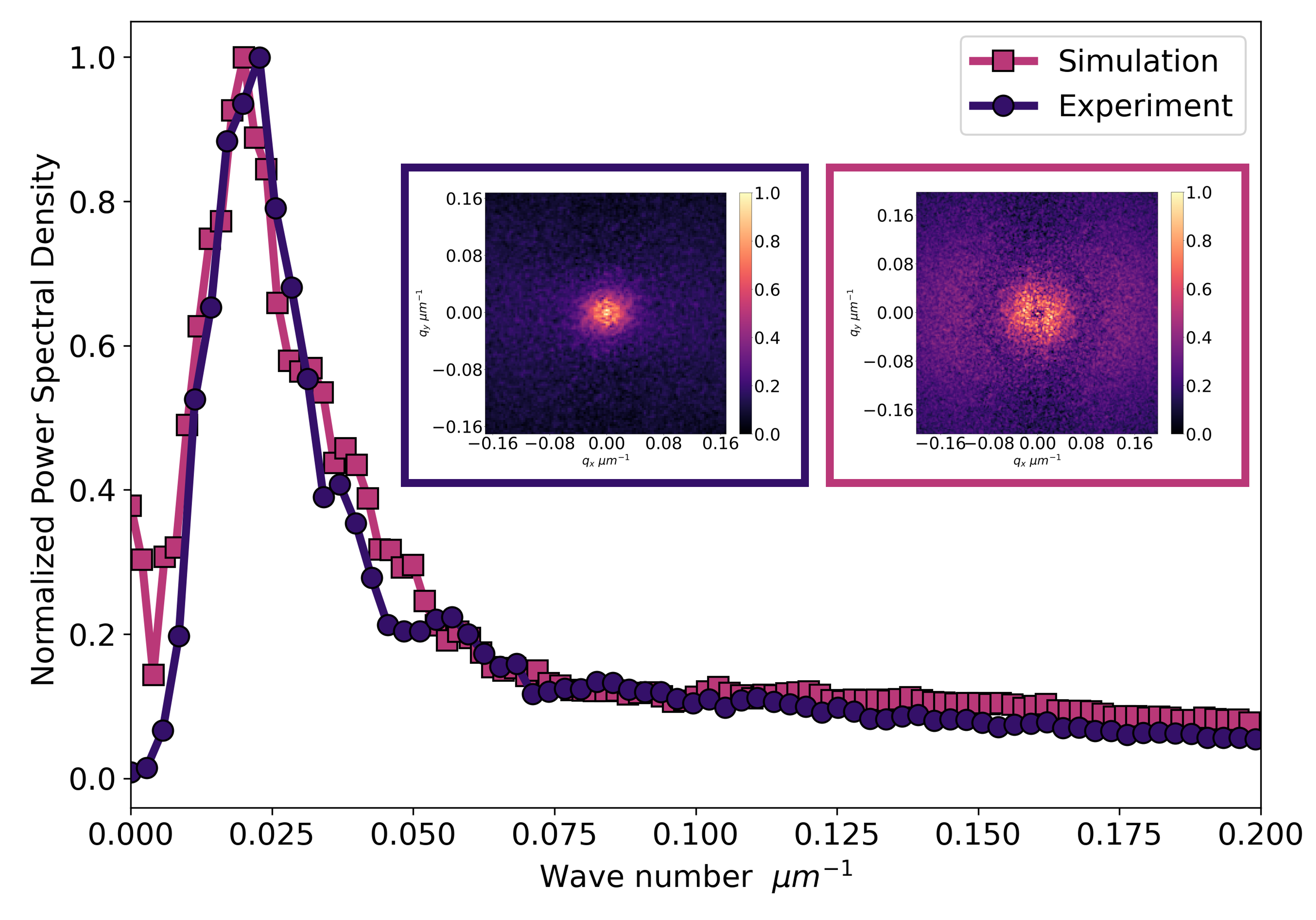}
        \caption{  The normalized power spectral density for the simulation and experiment. The steady state power spectrum is normalized by a power spectrum representative of the initial particle configuration. Peaks correspond to wavenumbers $ 0.020 \: \upmu m^{-1} $ ($50 \: \upmu m $)  and $ 0.023 \: \upmu m^{-1} $ ($ 43 \: \upmu m$), respectively. Insets show the absolute value of the Fourier amplitude spectra in log space (unnormalized) for (left) experiment and (right) simulation.} 
        \label{fig:ffts_w_experiment}
\end{figure}

To quantify how the pattern scale is coupled to $h^*$, we use spectral analysis of images to extract the pattern size as detailed in Methods, see Figure \ref{fig:ffts_w_experiment}. Radially averaging the 2D spectrum of the experimental and numerical patterns reveals a clear peak corresponding to the size scale of the pattern.
The pattern does not appear instantaneously, but emerges as a steady-state of large-scale density fluctuations after a short transient period. To ensure we measure the steady-state pattern, we characterize the transient timescale $t_{tr}$ and only examine the spectrum for $t > t_{tr}$. We average over many images ($\sim 50 $) due to the temporal fluctuations in pattern scale in our viewing window.  We find good agreement between the peak pattern size in experiments, 43 $\upmu$m, and in simulations (with lateral boundaries), 50 $\upmu$m.  

\begin{figure}[htbp]
    \centering
\includegraphics[width=\textwidth]{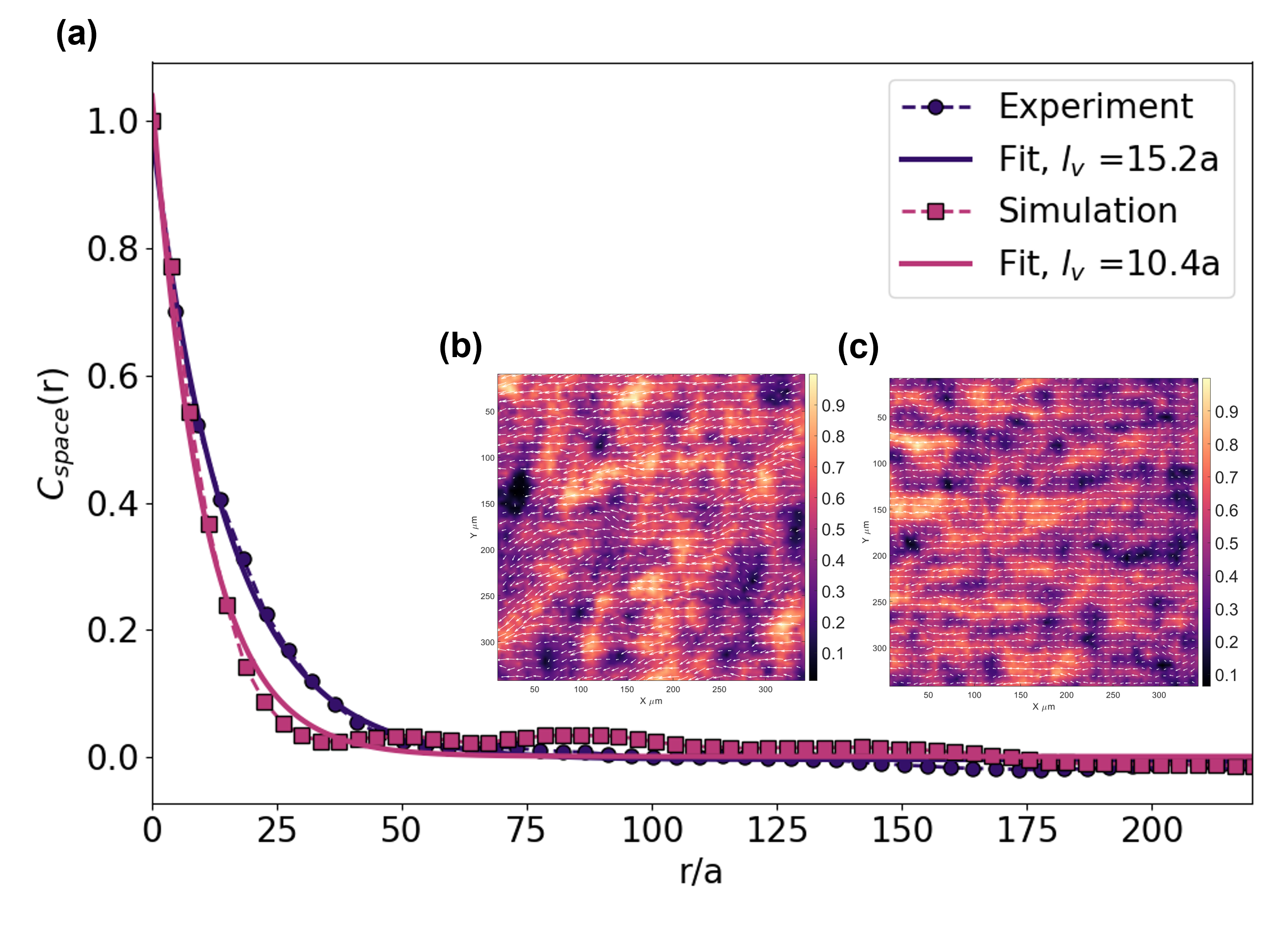}
        \caption{ Images:  (a) Spatial velocity autocorrelation function for the time averaged experimental and simulated velocity fields. Insets show the (b) Time averaged PIV velocity field of experimental images. (c) Time averaged velocity field of simulated images ($l^*=1000, h^*=10)$. Color-bar shows the normalized magnitude velocity.  }
        \label{fig:correlation_length}
\end{figure}

To characterize the patterning using a dynamical measurement we use spatial velocity correlations in the (time-averaged) suspension flow field. Specifically, we use the radial velocity auto-correlation function $C_{space}(r)$ described in Section~\ref{sec:correlation} to characterize the flow patterns observed in experiments and compare against simulation (see Figure~\ref{fig:correlation_length}). The decay of these velocity correlations is often used as a metric for pattern/cluster size in active and driven systems~\cite{snezhko2015velocity}.  Using an exponential fit, we extract the decay length $l_v$ from the $C_{space}(r)$ curves.  We good agreement between the experiments and simulations, with $l_v = 15.2 a$ in the experiments and $l_v = 10.4a$ in the simulations.  As this method uses flow information rather than particle positions, it serves as an alternate (and complimentary) metric with which to quantify pattern formation, and further validate that our large-scale simulations capture the density fluctuations observed in experiments.   In the following section, we characterize how these fluctuations are altered as a function of $h^*$.

\subsection{Pattern scale as a function of $h^*$.}

\begin{figure}[htbp]
    \centering
\includegraphics[width=\textwidth]{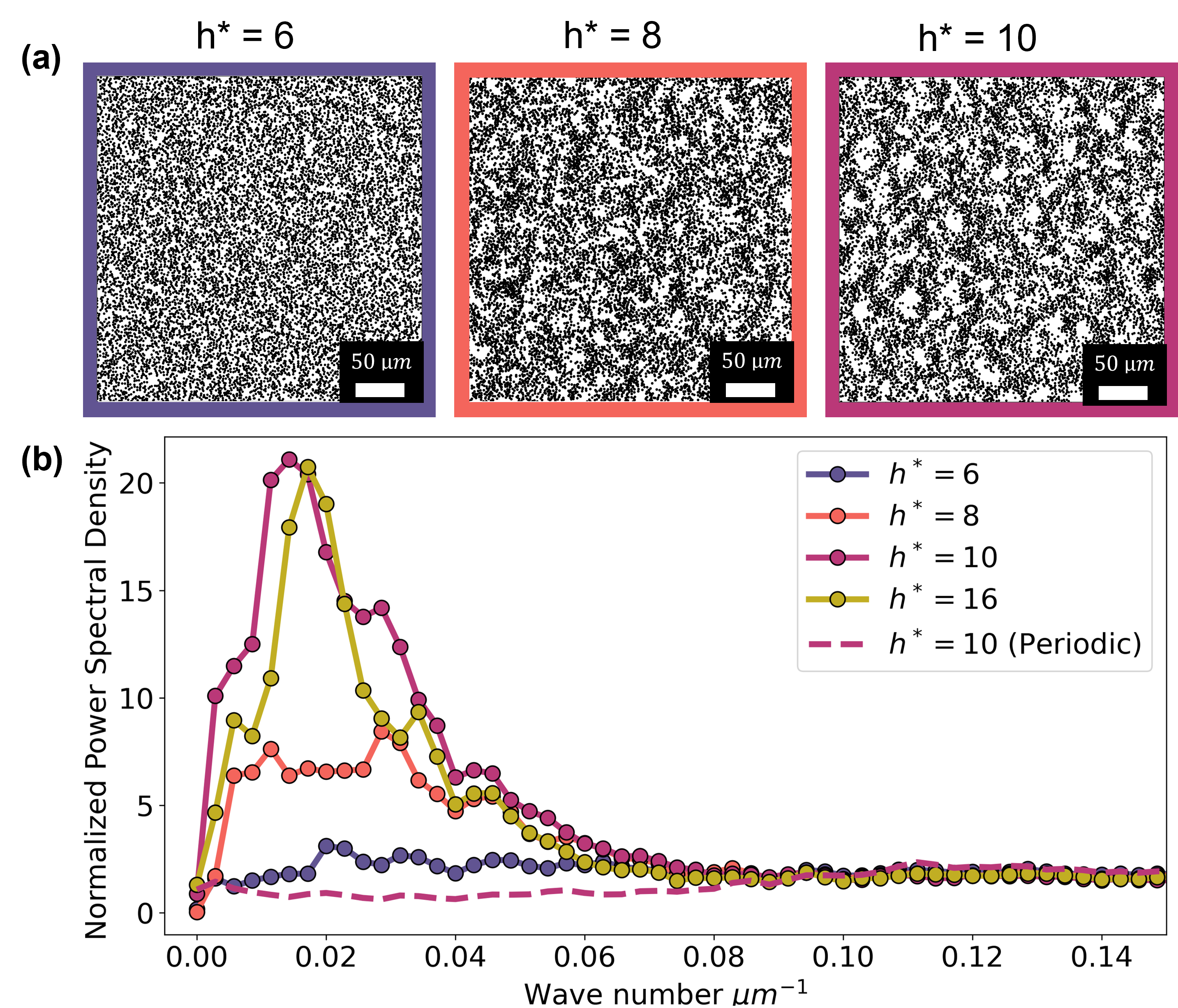}
        \caption{(a) Snapshots from simulation runs for $h^*=6,8,10$ (fixed boundaries, $l^*=500$) after a transient initialization period. (b) Normalized power spectral density for various $h^*$ (fixed walls), with a periodic simulation for $h^* = 10$ shown for reference.}
        \label{fig:simulation_ffts}
\end{figure}

To explore how changes in $h^*$ alter the patterning scale, we rely on our simulation measurements due to the challenges of experimentally fabricating a series of sample chambers with strong and varying confinement. We use a grid size of $l^*=500$ for this height study to reduce computational complexity, see SI Section IV for grid size comparison and SI Section III for details regarding computational run times. We find that the density fluctuations are determined by the height of the channel $h^*$ and are non-monotonic in $h^*$. At $h^*=10, 16$, the power spectral density shows that there is a clear peak in wavenumber (Figure \ref{fig:simulation_ffts}b) corresponding to the appearance of a large-scale pattern. Decreasing the height to $h^*=8$ makes the peak less apparent, and decreasing the height further to $h^*=6$ results in the peak disappearing entirely, e.g.\ we only observe particle-scale density fluctuations as seen in Figure \ref{fig:simulation_ffts}a). Our previous work demonstrates that at $h^* \gg 100$, the pattern again disappears and only particle-scale density fluctuations are observed~\cite{delmotte17}. This is expected, given the strong evidence we present here that the patterning is a result of vertical confinement.

\begin{figure}[htbp]
    \centering
\includegraphics[width=\textwidth]{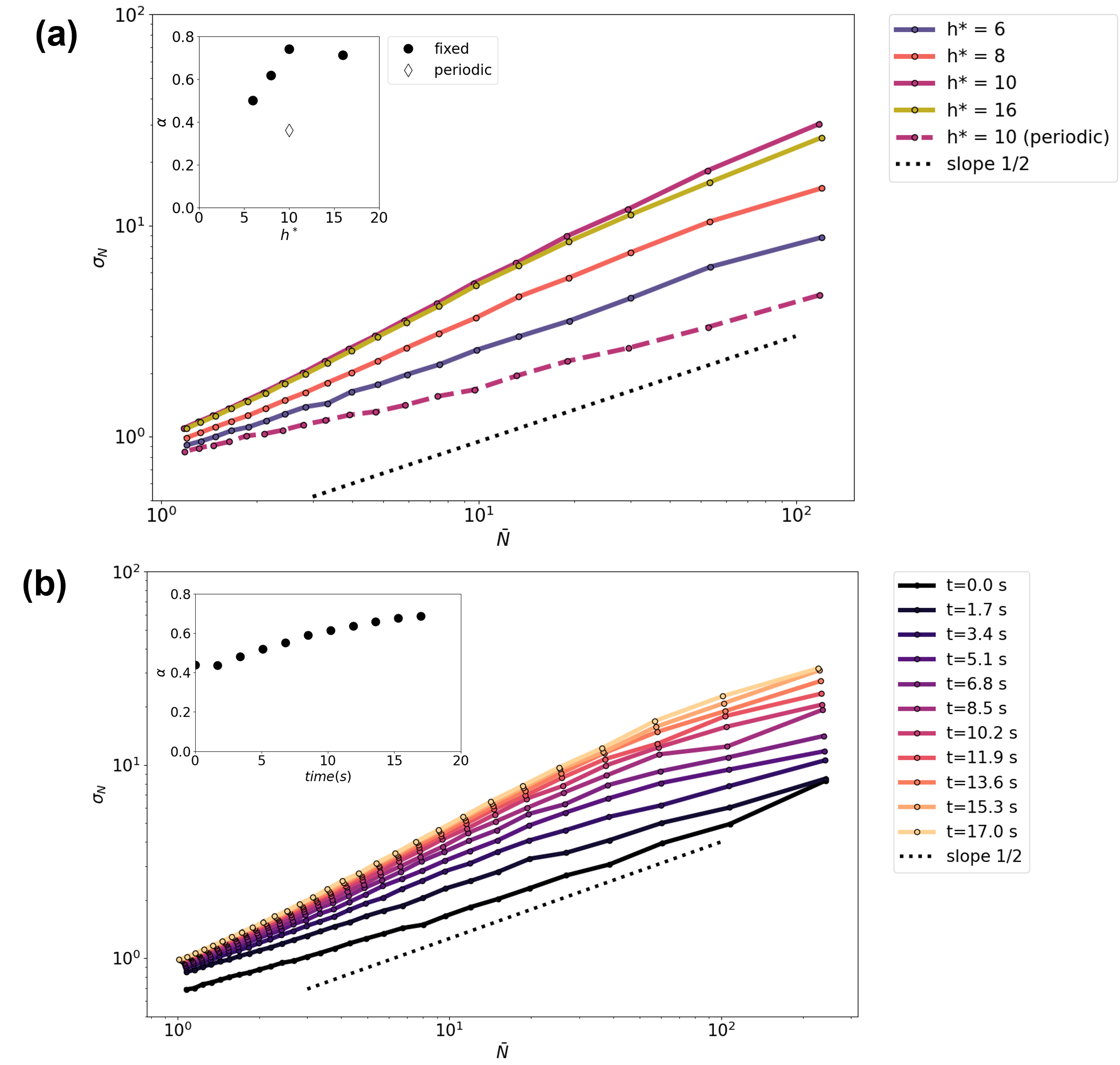}
        \caption{ Images:  (a) Standard deviation $\sigma(N)$ vs mean $\bar{N}$ plotted for different $h^*$, using the $l^*=500$ grid.  (b) Number fluctuations plotted at different instants in time using the $l^*=1000$ grid. }
        \label{fig:simulation_number_fluctuations}
\end{figure}

As an additional metric to quantify the density fluctuations we observe, we compute particle number fluctuations, a common metric for identifying clustering and other ordering~\cite{marchetti2013hydrodynamics}.  Dividing our system into boxes, we quantify how the standard deviation of particle number, $\sigma(N)$ varies as a function of mean particle number per box, $\bar{N}$ (e.g., box size).  In a random system, $\sigma(N) \sim \bar{N}^{1/2}$, whereas when ordering is present, $\sigma(N) \sim \bar{N}^\alpha$, where $\alpha > 1/2$.
We see that for $h^* = 10, 16$ where clustering is clearly observed,  $\sigma(N)$ grows as $\sim \bar{N}^{\alpha}$ with $\alpha = 0.7$ indicating strong clustering (see \cref{fig:simulation_number_fluctuations} a). This value also corresponds to the steady state $\alpha$ observed in \cref{fig:simulation_number_fluctuations} b.  The fluctuations are suppressed at $h^*=6$, where we find $\alpha = 0.5$,  the value expected for a two dimensional randomly distributed set of particles.  
Below, we discuss the mechanism which drives this non-monotonic dependence of pattern scale on $h^*$.

\section{Discussion}
\label{disc}



When the particles are suspended above a single boundary, we observe only particle-scale clusters that propagate dispersively~\cite{delmotte17}. When a microroller suspension is placed in strong vertical confinement ($h^* = 8, 10, 16$), we observe large-scale patterning (pattern size $\sim 50a$). Surprisingly, this patterning completely disappears when $h^* = 6$. To understand the origin of this non-monotonic effect, we turn to examining the flow fields generated by the microroller suspension.  Importantly, to see large-scale patterning, it was necessary to include lateral walls (at a distance $l^* \gg h^*$) in the simulations. These lateral walls are crucial to induce the density fluctuations, as the pattern is a result of large-scale flow recirculation that is unable to form when periodic boundary conditions are used.  We note that this recirculation can only alter suspension structure in a narrow range of $h^*$, as discussed in more detail below.

\begin{figure}[htbp]
    \centering
\includegraphics[width=0.9\textwidth]{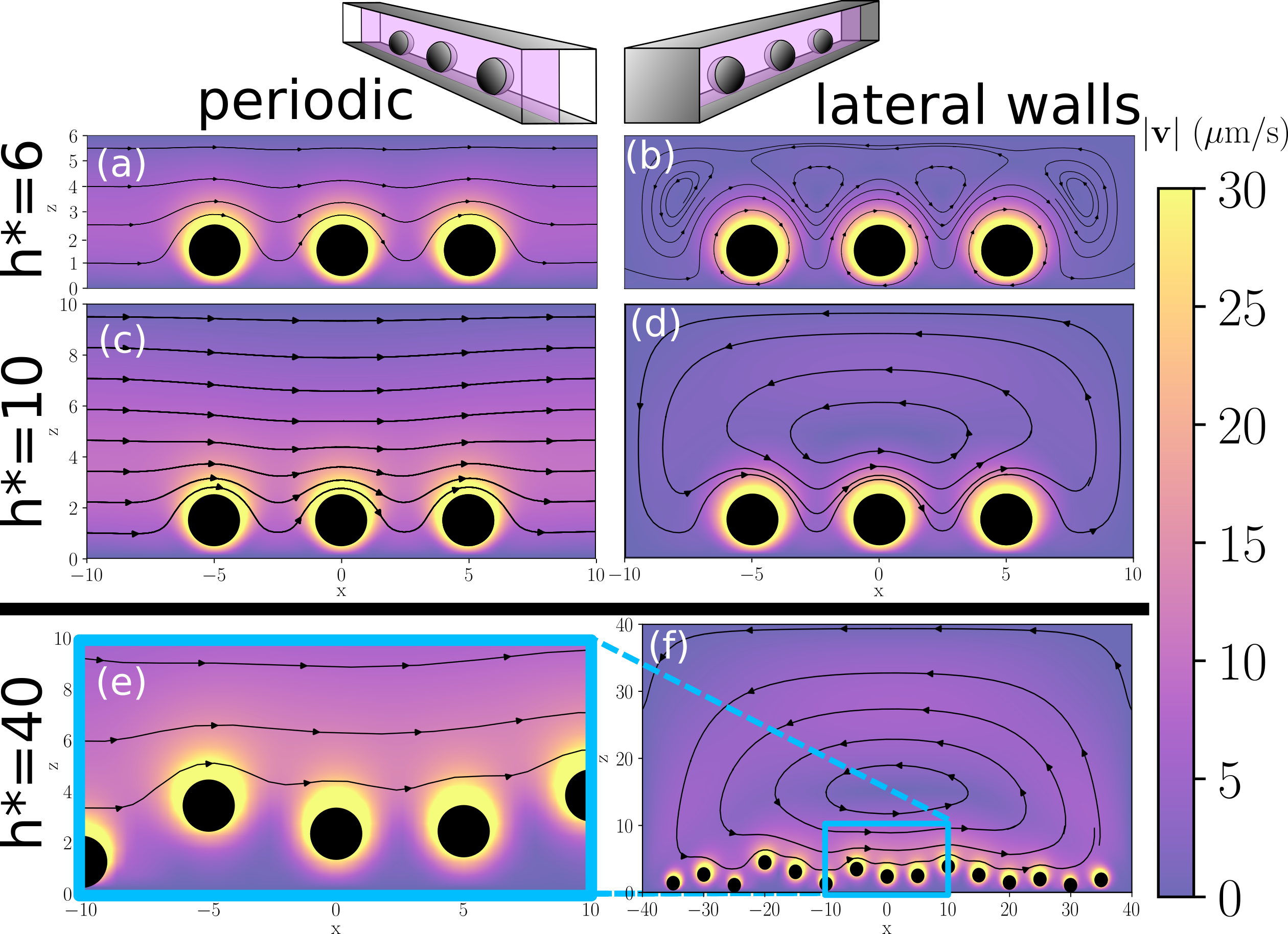}
        \caption{Visualizations of 2D slices from simulated 3D velocity fields around a small suspension of microrollers. Periodic simulations (a,c) are unconfined in the plane, while lateral wall simulations (b,d,f) have fixed walls at both ends of the domain in x. The particle heights in (e,f) are sampled from simulation data. All figures show the full computational domain except (e), which is a zoom-in from (f).}
        \label{fig:streamlines}
\end{figure}

Figure \ref{fig:streamlines} illustrates how key features in the microroller-generated flow field evolve as a function of $h^*$.  To visualize the flow-field created by a suspension of microrollers, we discretize each sphere using an micosphere with 642 vertices (blobs) and use the rigid multiblob method to more accurately resolve the flow-field~\cite{rigid_multiblob, rigid_multiblob_review}.  The streamlines show that at a confinement of $h^*=6$, lateral walls induce a complex flow, but no large-scale recirculation. When $h^* = 10$, a system-spanning recirculation zone develops.  This same recirculation zone is present when $h^* = 40$, but the recirculation center is far above the rollers and locally (e.g., to the rollers) the field appears vortex-free. We note that the particle heights in Figure \ref{fig:streamlines} (e) and (f) are sampled from simulation data, and show that out-of-plane fluctuations in particle position do not disrupt the overall recirculation.  Since a large-scale simulation using $h^*=40$ is computationally infeasible, we sample particle heights from a simulation using $h^*=16$ after verifying the same height distribution is obtained in a simulation with a smaller channel height of $h^*=10$. Our flow-field calculations illustrate that the presence of the recirculation center is strongly implicated in triggering the large-scale pattern formation, but its center needs to be close enough to the particles for them to interact with it.  

Further evidence for this recirculation as the patterning mechanism is provided by the flow field at $h^*=6$:  the large-scale recirculation vanishes, and breaks into many particle-scale recirculation zones.  Thus, while the flow field is qualitatively different in all three cases, it is only when a recirculation zone with a center near to the particle size is present that large-scale density fluctuations appear in the driven suspension.  We hypothesize that this particle-scale recirculation zone is necessary so that particle can sample the `backwards' flows created on the chamber ceiling.  These flows then result in strong heterogeneity in the suspension (see Figure~\ref{fig:velocity_distributions}b), and appear to amplify the ever-present particle-scale density fluctuations~\cite{delmotte17} into large-scale structures.

\section{Conclusions and Outlook}

We have identified an unexpected consequence of vertically confining a rotationally-driven suspension: large-scale ($\sim50a$) patterns appear.  This effect is linked to the long-range lateral flowfield generated by the microrollers, and as we numerically demonstrate, distant lateral walls are required to initiate the emergence of the patterning.  We further show that the role of the lateral walls is to create a recirculation zone; this recirculation only leads to large-scale density fluctuations when the height of its center is comparable to the particle size.  This particle-scale recirculation zone leads to strong mixing and heterogeneity in the suspension, leading to persistent but transient density fluctuations.  
While we have identified the general mechanism which initiates density fluctuations, it remains an open question what sets their particular length scale, and whether this length scale is coupled to suspension density.

Understanding how confinement modifies suspension transport is crucial for many biomedical applications, such as cell sorting, targeted drug delivery, and advanced diagnostics~\cite{zhang2024nanorobot,li2017micro, chen2024hollow, yu2019characterizing,wu_medical_2020}.  As microrollers have been demonstrated to show potential in many of these applications~\cite{chen2021overview}, careful consideration of the strong effects of confinement on their suspension dynamics is warranted.  Moreover, perhaps with a deeper mechanistic understand of confined-induced emergent structure formation, this effect can be leveraged to expand the biomedical applications prospects for  microroller-focused use cases.

\section{Supplementary Material}

The supplementary material contains complete list of simulation parameters, details regarding transient time characterization, computational grid design, simulation wall times, details of grid size study and details describing the supplementary videos.

\section{Acknowledgments}
We thank Chris Jacobsen for discussion regarding spectral analysis, and Blaise Delmotte for numerous
informative discussions regarding the recirculation mechanism driving pattern formation.
 This work was partially supported by
the National Science Foundation under award number DMR-2011854. This research was supported in part through the computational resources and staff contributions provided for the Quest high performance computing facility at Northwestern University which is jointly supported by the Office of the Provost, the Office for Research, and Northwestern University Information Technology.

\newpage

\clearpage

\newpage

\bibliographystyle{unsrt}
\bibliography{references}

\end{document}